# Static Virus Spread Algorithm for DNA Sequence Design

Yao Yao, Xun Zhang, Xin Liu, Yuan Liu, Xiaokang Zhang, Qiang Zhang

*Abstract*—DNA is not only the genetic material of life, but also a favorable material for a new computing model. Various research works based on DNA computing have been carried out in recent years. DNA sequence design is the foundation of such research. The sequence quality directly affects the universality, robustness, and stability of DNA computing. How to design DNA sequences depends on the biological properties and target requirements, which is a typical combinatorial optimization problem. In this paper, in order to design DNA sequences with high-quality, we propose a novel meta-heuristic evolutionary algorithm, termed the static virus spread algorithm (SVS). Through this algorithm, we focus on the constraints of universal DNA sequence design and produce a large number of DNA sequences with non-complementarity and small difference in melting temperature as the objectives, and fully considering the balanced proportion of the four bases. The computer simulation and polyacrylamide gel electrophoresis experiments show that the high-quality DNA sequences designed by this algorithm are effective, which is expected to provide a convenient tool for sequence preparation before DNA biochemical operations.

This work was supported by the National Key Technology R&D Program of China (No. 2018YFC0910500), the National Natural Science Foundation of China (Nos. 62272079, 61751203, 61972266, and 61802040), LiaoNing Revitalization Talents Program (No. XLYC2008017), the Innovation and Entrepreneurship Team of Dalian University (No. XQN202008), and the Natural Science Foundation of Liaoning Province (Nos. 2021-MS-344, 2021-KF-11-03, and 2022-KF-12-14). (Corresponding author: Qiang Zhang.)

Yao Yao, Xun Zhang, Xin Liu, Yuan Liu, Xiaokang Zhang are with the School of Computer Science and Technology, Dalian University of Technology, Dalian, Liaoning 116024, China (e-mail: yy@mail.dlut.edu.cn; madao@mail.dlut.edu.cn; xinliuaisky@mail.dlut.edu.cn; liuyuan.dlut@gmail.com; xiaokangzhangdl@gmail.com).

Qiang Zhang was with the Key Laboratory of Advanced Design and Intelligent Computing, Ministry of Education, School of Software Engineering, Dalian University, Dalian 116622, China. He is now with the School of Computer Science and Technology, Dalian University of Technology, Dalian 116024, China (e-mail: zhangq@dlut.edu.cn).

## I. INTRODUCTION

DNA computing is a new computing technique that encodes information into DNA molecules and then operates on DNA molecules to achieve computing functions. Since Leonard M. Adleman established the first solution to the Hamilton path problem based on DNA molecules in 1994 [1], scientists have carried out a lot of research on biological computing models [2][3] and biological nanostructures [4][5][6][7] and related functions. With the development of computational theories and biotechnology, molecular logic circuits [8][9], artificial biological neurons [10][11], artificial intelligence based on biological materials [12][13][14], efficient encoding/decoding of DNA information [15][16][17], storage technologies [18], and other fields have emerged, leading to a new stage of cross-convergence and development of computer and biological nanotechnology. However, the abovementioned research works need to optimize the design of DNA sequences to ensure the reliability of biochemical operations. DNA strands present complex attractive and repulsive states in the reaction solution. Problems such as false-positive pairing, abnormal secondary structure, and high similarity [19] between DNA strands will directly affect the efficiency and process of biochemical reactions. Therefore, how to efficiently obtain the ideal DNA sequence set has become the focus of many researchers.

DNA sequence design is a complex combinatorial optimization problem, which not only has diverse design objectives for diverse experimental requirements, but also is affected by the biological properties of DNA, and required to satisfy multiple constraint conditions. In the original phase of DNA sequence design, due to insufficient theoretical basis, researchers mainly used a variety of methods to randomly generate sequences [20][21][22][23], and search for lower similarity sequences. The screening workload was huge, the efficiency was low, and the contingency was strong. In the subsequent research, with further exploration of DNA properties, some scientists began to design sequences based on the biochemical principles of DNA. It has gradually become a combinatorial optimization problem with thermodynamic properties, similarity, and stability as the objectives, with secondary structure, base continuity, Guanine-Cytosine (GC) content, and Hamming distance as the constraint conditions. A variety of heuristic algorithms to design DNA sequence began to appear. For the Hamming distance, researchers have used an algorithm based on minimum free energy (MFE) to reduce the occurrence of DNA mismatches and secondary structures [24].

On this basis, they further optimized the precocity of genetic algorithm and used it for the design of DNA sequences. Compared with the genetic algorithm before improvement, the new method effectively improved the sequence stability [25]. The weed algorithm based on niche crowding [26], chaotic whale algorithm [27], and whale search algorithm [28] proposed a new theory and constraints of DNA sequences and obtained a better Hamming distance. For the thermodynamic or stability, researchers proposed a NACST system based on a multi-objective evolutionary algorithm [29], a dynamic adjustment strategy for the Bloch quantum chaos algorithm [30], a non-dominated sorting method combining the characteristics of the bat algorithm and particle swarm algorithm [31], a dynamic membrane evolutionary algorithm combining membrane activity and segmentation rules [32], and a bacterial foraging algorithm based on a competition mechanism [33]. These methods combined multiple algorithms (mechanisms) to design DNA sequence sets with thermodynamic stability and structural stability. In addition, in order to obtain DNA sequence sets efficiently, a dynamic membrane algorithm combining an adaptive differential evolution algorithm and a particle swarm optimization algorithm effectively increased the update efficiency and global search ability and had higher efficiency in obtaining DNA sequences [34]. Or the genetic algorithm can be optimized by introducing a variety of constraints in the process of non-dominated sorting to obtain good population diversity and algorithm convergence, which was very helpful to quickly obtain DNA sequences [35]. Researchers also proposed a particle swarm optimization algorithm combined with an elastic collision mechanism that enhanced the search ability and improved the efficiency of obtaining DNA sequences [36]. These algorithms can quickly filter the reasonable search space and then quickly extract the target DNA sequences.

In this paper, in order to obtain high-quality DNA sequence sets with high efficiency, inspired by previous work, we proposed a static virus spread algorithm (SVS) to design DNA sequences with seven constraints, namely Similarity, H-measure, GC Content, Continuity, Hairpin Structure, Melting Temperature and Self-Dimer. With the continuous iteration of the algorithm, high-quality DNA sequences with low complementarity probability, no hairpins, low continuity, and little change in melting temperature can be generated. Finally, the effectiveness of this algorithm was proved by computer simulation and polyacrylamide gel electrophoresis experiments.

The research significance is reflected in two aspects:

1) The DNA sequence designed in this article not only meets the requirements of similarity and melting temperature in the past, but also comprehensively considers the proportion balance of the four bases and significantly improves the non-complementarity between sequences.

2) We simulated the propagation process of viruses in a static environment and proposed an SVS algorithm with strong convergence ability. Using this algorithm, we achieved the design of DNA sequences.

The rest of this article is arranged as follows: The second part introduces the theoretical basis of DNA sequence design and summarizes numerous constraint models. In the third part, we introduce the SVS algorithm model and the detailed steps of using it to design DNA sequences. The forth part presents the results of the algorithm, through computer simulation and gel electrophoresis experiments to prove that the DNA sequences designed by the SVS algorithm have good non-complementarity and low differential melting temperature characteristics. The final part summarizes the article and suggests areas for future research.

## II. DNA Sequence Design Model

The DNA double-helix structure of specific pairing of four bases was proposed in the Watson–Crick model [37], which made DNA research enter a new stage. With the in-depth study of DNA structure, scientists have found many characteristics of DNA, such as the melting temperature of the double-helix structure, the self-reverse folding of a single strand, etc. On this basis, many DNA sequence design theories have been proposed. For DNA computing, the use of DNA strands to form various structural units or computing modules must ensure that the required DNA strands can only hybridize in accordance with the expected binding domain; otherwise, there will be multiple reaction leaks or other uncontrollable situations. Therefore, designing DNA sequences with non-complementarity is particularly important. We take Similarity and H-measure between two DNA sequences as the main constraint conditions to fully reduce the possibility of similarity and complementarity of DNA strands. In addition, the thermodynamic properties of DNA strands are also considered when conducting biochemical reactions, because they affect the temperature control in the annealing process. Therefore, ensuring that the difference between the Melting Temperatures is small is also a constraint we focus on. For other constraints of the DNA strand itself, such as Continuity, GC base content, Hairpin Structure and Self-Dimer, we consider them as secondary conditions.

### A. Similarity

The magnitude of the Similarity measure can help determine the similarity between two DNA sequences. The larger the Similarity, the more similar the two DNA sequences. For a DNA sequence set $X$, there are $m$ sequences. The similarity of sequence $X_i$ is calculated by comparing the sliding Hamming distance between $X_i$ and the other $m-1$ sequences [38], which is expressed as (1):

$$f_{Similarity} = \sum_{i=1}^{m} \sum_{j=1, i \neq j}^{m} Similarity(X_i, X_j) \quad (1)$$

where $X_i$ and $X_j$ are parallel sequences in set $X$. $Similarity(X_i, X_j)$ is divided into two parts for calculation as shown in (2), termed discontinuity-similarity (SDis) and continuity-similarity (SCon). The maximum similarity value generated by two parallel sequences during sliding is calculated. The $Shift$ function is used to represent the sliding process. The $g$ gaps generated during sliding are supplemented by (-). $0 \leq g \leq n-1$, and $n$ is the total number of bases in the DNA sequence. The sliding distance is expressed by $k$, and $-n \leq k \leq n$:

$$Similarity(x,y) = \max_{g,k}(SDis(x, Shift(y(-)^g\ y, k)) \\ + SCon(x, Shift(y(-)^g\ y, k))) \quad (2)$$

Equations (3) and (4) represent discontinuity-similarity and continuity-similarity, respectively:

$$SDis(x,y) = T(\sum_{i=1}^{n} eq(x_i, y_i), DS \times n),\ DS \in [0,1] \quad (3)$$

$$SCon(x,y) = \sum_{i=1}^{n} T(subeq(x,y,i), CS),\ CS \in [1,n] \quad (4)$$

where $DS = 0.17$, and $CS = 6$ [26]. Equations (5)-(7) describe the functions in (3) and (4), and $subeq(a,b,i)$ is used to obtain the number of consecutive identical bases from the $i^{th}$ position:

$$T(a,b) = \begin{cases} a & a > b \\ 0 & a \leq b \end{cases} \quad (5)$$

$$eq(a,b) = \begin{cases} 1 & a = b \\ 0 & a \neq b \end{cases} \quad (6)$$

$$subeq(a,b,i) = \begin{cases} 1 & a_i = b_i\ \&\&\ a_{i+1} = b_{i+1} \\ 0 & otherwise \end{cases} \quad (7)$$

*B. H-measure*

In addition to Similarity, the H-measure is used to evaluate the complementarities between the DNA sequence $X_i$ (5'-3') and the reverse sequences (3'-5') of other DNA sequences in the set. Refer to the calculation method in [38], which is expressed as (8):

$$f_{H-measure} = \sum_{i=1}^{m}\sum_{j=1, i \neq j}^{m} H-measure(X_i, X_j) \quad (8)$$

It is divided into the discontinuity H-measure (HDis) and continuity H-measure (HCon), expressed as (9)-(11):

$$H-measure(x,y) = \max_{g,k}(HDis(x, Shift(y(-)^g\ y, k)) \\ + HCon(x, Shift(y(-)^g\ y, k))) \quad (9)$$

$$HDis(x,y) = T(\sum_{i=1}^{n} cb(x_i, y_i), DH \times n),\ DH \in [0,1] \quad (10)$$

$$HCon(x,y) = \sum_{i=1}^{n} T(subcb(x,y,i), CH),\ CH \in [1,n] \quad (11)$$

where $DH = 0.17$, and $CH = 6$ [26]. The parameters in the H-measure calculation are basically the same as those in the measurement of Similarity. $subcb(a,b,i)$ is used to obtain the number of consecutive complementary bases from the $i^{th}$ position:

$$cb(a,b) = \begin{cases} 1 & a = b^{rc} \\ 0 & a \neq b^{rc} \end{cases} \quad (12)$$

$$subcb(a,b,i) = \begin{cases} 1 & a_i = b_i^{rc}\ \&\&\ a_{i+1} = b_{i+1}^{rc} \\ 0 & otherwise \end{cases} \quad (13)$$

where $b^{rc}$ is the reverse complement sequence of $b$, for instance, $b = 5'-x_1 x_2 x_3 \ldots x_n-3'$, the reverse sequence $b^r = 3'-x_n \ldots x_3 x_2 x_1-5'$, and $b^{rc} = 3'-\bar{x}_n \ldots \bar{x}_3 \bar{x}_2 \bar{x}_1-5'$.

*C. GC Content*

GC pair content is the percentage of the bases G (Guanine) and C (Cytosine) in a DNA sequence. Among the four pairs of DNA bases, A(Adenine) = T (Thymine) is paired with two hydrogen bonds, whereas $G \equiv C$ is paired with three hydrogen bonds. Therefore, $G \equiv C$ has a greater impact on the structural stability and the melting temperature of double-stranded DNA (dsDNA). The content is usually about 50%. GC content can be expressed as (14):

$$GC = \frac{sum('G') + sum('C')}{n} \quad (14)$$

where $sum('G')$ and $sum('C')$ represent the total number of G and C bases in each sequence, respectively.

*D. Continuity*

The Continuity constraint refers to the continuous occurrence of multiple identical bases [38]. Excessive consecutive identical bases may cause abnormalities in the secondary structure (especially Guanine), which should be avoided in conventional DNA sequence design.

It should be noted that Continuity is not a hard constraint in the actual DNA sequence design. In some cases, it is necessary to design or add multiple continuous bases to obtain specific structural connectivity or to distinguish products. For instance, in the gel electrophoresis experiment, if the length of the DNA product strand and the input strand is very close or the same, this cannot be distinguished in the gel. In this case, several consecutive bases (usually Thymine) should be added behind the non-reaction domain of the DNA input strand or product strand, so that they can be distinguished on the gel.

*E. Hairpin Structure*

As the name implies, Hairpin Structure evaluation is used to judge whether single-stranded DNA (ssDNA) may form a hairpin structure [38].

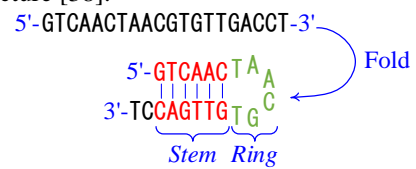

Fig. 1. Example of the Hairpin Structure

Hairpin structure refers to the bending and folding of ssDNA to itself to form a structure similar to a hairpin, as shown in Fig. 1. There are usually two conditions to form a hairpin structure:

1) There is a hairpin ring large enough to ensure that the DNA strand can be bent back.

2) On both sides of the hairpin ring, there is a complementary palindrome sequence that is long enough to ensure that the hairpin has a sufficiently stable stem.

The calculation method can be expressed as (15) and (16):

$$Hair = \sum_{P=P_{min}}^{(n-R_{min})/2} \sum_{r=R_{min}}^{n-2p} \sum_{i=1}^{n-2p-r} T(\sum_{j=1}^{pri} cb(X_{i+j}, X_{n-j}), \frac{pri}{2}) \quad (15)$$

$$pri = min(p+i, n-p-i-r) \quad (16)$$

where $r$ is the length of ring, and $p$ is the length of the stem. $R_{min}$ and $P_{min}$ are the minimum length ring and stem that can form a hairpin, respectively. In the actual calculation, $R_{min}=6$, and $P_{min}=6$ [26].

### F. Melting Temperature

The Melting Temperature (Tm) is used to describe the temperature when 50% of dsDNA opens to form ssDNA. The double-helix structure of DNA is a stable structure, which will gradually become single strands with the increase in temperature and will recover with the decrease in temperature. The whole process is called DNA annealing. In the biochemical operations of DNA, multiple DNA strands will be put into the same Eppendorf (EP) tube and annealed to form the target substrate. Therefore, Tm is an important constraint. The smaller the Tm difference between DNA strands, the more conducive to temperature control. The Tm value is measured according to the free energy model in [35].

### G. Self-Dimer

The Self-Dimer (SD) constraint, which is also a constraint based on the Hamming distance, is a supplement to the H-measure. All the DNA sequences we designed were in the direction of 5'-3'. For DNA sequence $X_i$ (direction 5'-3'), its reverse sequence is $X_i^r$ (direction 3'-5'). When $X_i$ and $X_i^r$ partially complement each other, they are termed Self-Dimer (Fig. 2).

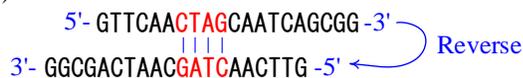

Fig. 2. Example of the Self-Dimer

This is usually to be avoided in the biochemical reactions of DNA computing. Under this circumstance, the DNA $X_i$ may mask the key recognition domain that needs to participate in the biochemical reaction and reduce the concentration of $X_i$ in the reaction system, which is very unfavorable to the biochemical reaction. The judgment method is based on the sliding Hamming distance, and expressed in (17) and (18):

$$SD = max_{-n<k<n}(n-|k|-Ham(X_i, Shift(X_i^{rc}, k))) \quad (17)$$

$$Ham(a,b) = \begin{cases} 1 & a \neq b \\ 0 & a = b \end{cases} \quad (18)$$

where $X_i^{rc}$ is the reverse complementary sequence of $X_i$, and $Ham(a,b)$ is used to calculate the Hamming distance. The DNA sequence designed based on this constraint can largely avoid the occurrence of SD, and thus effectively improve the reliability of the DNA biochemical reactions and reduce the reaction leakage.

## III. STATIC-VIRUS SPREAD ALGORITHM

Virus have strong reproductive capacity in the process of spread and mutation has not only attracted the research of scholars in biology and medicine, but also attracted the attention of scholars in the computer field for its spreading mechanism [39]. The virus spread algorithm, like many heuristic bionic algorithms, has a strong optimization ability and a more efficient iteration speed. It has a strong ability to solve many combinatorial optimization problems [40][41][42].

Inspired by the virus spread model [39], and combined with the epidemic prevention method of standstill orders in China, we propose a novel meta-heuristic optimization algorithm, namely the static virus spread (SVS) algorithm. This algorithm takes the relatively fixed position of the hosts as the premise and endows the virus with the non-contact spread mechanism of aerosols. The design of DNA sequences based on the SVS algorithm is to represent DNA sequences as viruses. At the initial stage, viruses spread to a certain population, and the optimization of the DNA sequence is realized through the evolution of the viruses in the infected hosts. The amplification of the DNA sequence set can be realized through the process of continuous spread of the infected hosts to the surrounding susceptible population. With the continuous optimization of DNA, the virus is replicating and evolving toward higher spread capacity and lower toxicity. The cure or death mechanism of the infected host is used to avoid the situation of falling into the local optimum. By gradually establishing the immunologic barrier, the search space is reduced and the convergence of the algorithm is improved, thus providing the algorithm with high operating efficiency.

### A. Overview of the Static-Virus Spread Algorithm Model

The viruses mentioned in this article have the ability to spread in the air. It is stipulated that during the epidemic period of the virus, there is no treatment, and the infected hosts will eventually die or self-cure. The spread mechanism is shown in Fig. 3.

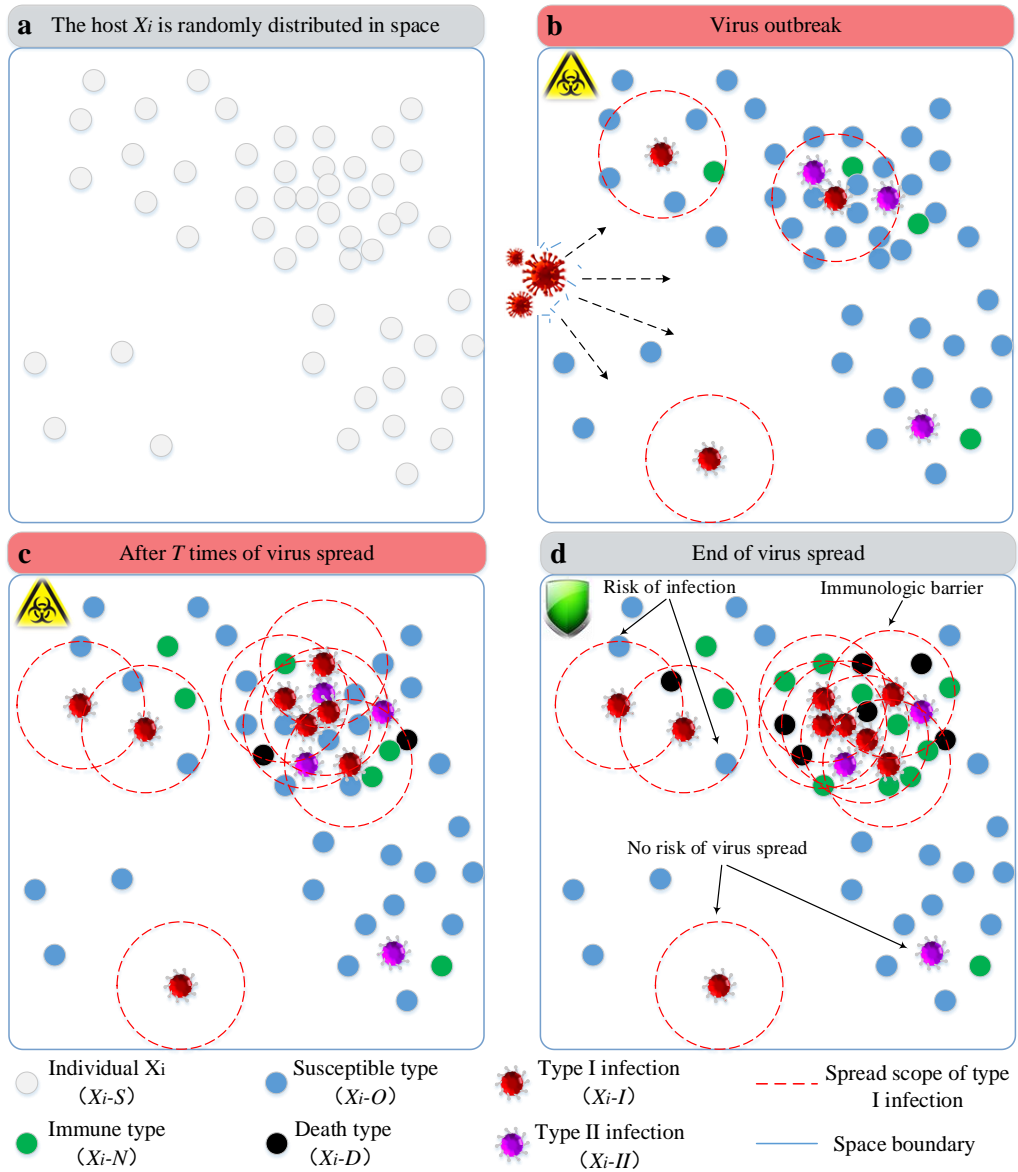

Fig. 3. Schematic Diagram of the Static-Virus Spread Algorithm

In Fig. 3, there are six balls and two lines with different colors, whose meanings are as follows:

1) The grey ball is abbreviated as $X_i - S$. Each ball represents a host $X_i$. This algorithm is only used to mark the initial position of $X_i$.

2) The blue ball is abbreviated as $X_i - O$, it indicates that $X_i$ is a susceptible type, and it can be infected by viruses.

3) The red ball is abbreviated as $X_i - I$, it indicates that $X_i$ is a type I infection host. It has the ability to continue to spread viruses around. The final result is self-cure or death.

4) The magenta ball is abbreviated as $X_i - II$, it indicates that $X_i$ is a type II infection host. It will not spread the virus. The final result is self-cure or transforming into type I infection with a certain probability.

5) The green ball is abbreviated as $X_i - N$, it indicates that $X_i$ has immunity and will not be infected.

6) The black ball is abbreviated as $X_i - D$, it indicates that $X_i$ was infected but is now dead.

7) The red dotted line indicates the effective spread range of type I infection hosts. All susceptible individuals within this range are likely to be infected.

8) The solid blue line indicates the boundary of all $X_i$ living areas.

It should be noted that $X_i - I$ and $X_i - II$ will transform into $X_i - O$ or $X_i - N$ with a certain probability after self-cure.

The algorithm consists of four stages:

The first stage is initialization. A group of hosts ($X_i$) live in the 2D space ($Lm \times Ln$), and their positions are fixed from beginning to end. The initialized position information is shown in the gray ball in Fig. 3a, and their distribution positions are sparse or dense. At this moment, there is no virus intrusion.

In the second stage, there is a virus outbreak. As shown in Fig. 3b, a small number of viruses invade this space, and some of $X_i - O$ are infected and show different infection types (Type I or Type II). Some lucky $X_i$ have natural immunity.

The third stage is the epidemic period of the virus. During this period, $X_i - I$ continues to spread virus to $X_i - O$ in its effective spread distance, and the number of infected $X_i$ increases sharply. In the process of virus infection and spread, the virus continues to evolve. In the following time, $X_i - I$ and $X_i - II$ are constantly transformed into $X_i - O$ and $X_i - N$ through self-cure, or $X_i - I$ can be transformed into $X_i - D$ by death; and some $X_i - II$ are transformed into $X_i - I$ with a certain probability. The situation shown in Fig. 3c is gradually formed.

The fourth stage is the end of the virus epidemic. There are two end marks, as shown in Fig. 3d:

1) The algorithm has reached the specified epidemic time, even though it still has spread risk.

2) The number of $X_i - N$ reaches a certain threshold, thus forming an immunologic barrier.

When the virus epidemic is over, we can screen out the virus set $V$ carried by $X_i - I$ and $X_i - II$. It means that the optimized virus library has been obtained (high spread capacity means high quality).

### B. DNA Sequence Design Based on the SVS Algorithm

We define each virus $V_i$ as a DNA sequence, and each host $X_i$ may become a vector of $V_i$. $V_i$ spreads and evolves among $X_i$. Finally, a set of DNA sequences meeting the low similarity and complementarity probability constraints are obtained through screening. Fig. 4 shows the main flowchart of the algorithm.

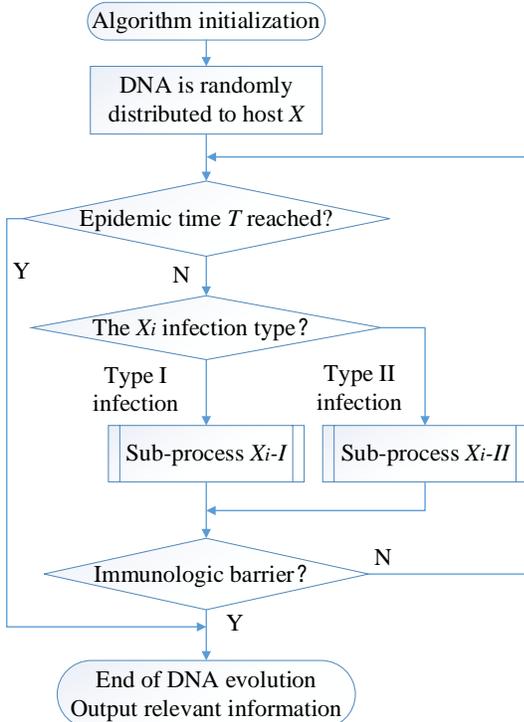

Fig. 4. Flowchart of the Main Algorithm

The main algorithm is divided into seven steps:

Step 1. Parameter initialization. The initial number of hosts $X_i$ and DNA $V_i$ is $h$ and $m$, respectively, and the location information $Loc$ of each $X_i$, the space range $Lm \times Ln$, and the immunity of $X_i$ are initialized. The location information can be expressed as (19):

$$Loc(X_i) = (lm, ln) \quad lm \in [0, Lm], \quad ln \in [0, Ln] \quad (19)$$

Step 2. Spread of DNA among $X$. Some $X_i - O$ will be infected. They will be the source of this round of epidemics. $X_i - I - V_j$ indicates that $X_i$ is a type I infection, and the DNA number carried is $V_j$. We define all the initial infections as type I infections, $i \in [0, h]$, $j \in [0, m]$, $i, j$ is unique:

$$(X_i - O) \rightarrow (X_i - I - V_j) \quad (20)$$

Step 3. If the maximum number $T$ of virus spread has not been reached, step 4 will be carried out; otherwise, the last step will be carried out.

Step 4. Determined the infection type of all $X_i$ one by one. If it is $X_i - I$, then go to sub-step 5.1. If it is $X_i - II$, then go to sub-step 5.2.

Steps 5.1 and 5.2. As these two steps are complex, they are separated into sub-processes, which will be described in detail later.

Step 6. Determined whether the immunologic barrier is established. If the number of $X_i - N$ reaches the corresponding scale, the virus epidemic will be ended in advance, otherwise, return to step 3.

Step 7. After the virus epidemic ends, we will screen out $V_i$ we need as the final DNA sequence from the DNA library $V$.

Fig. 5 shows a sub-process of the main algorithm, which is the operation for $X_i - I$ during the virus epidemic period corresponding to step 5.1 in the main algorithm.

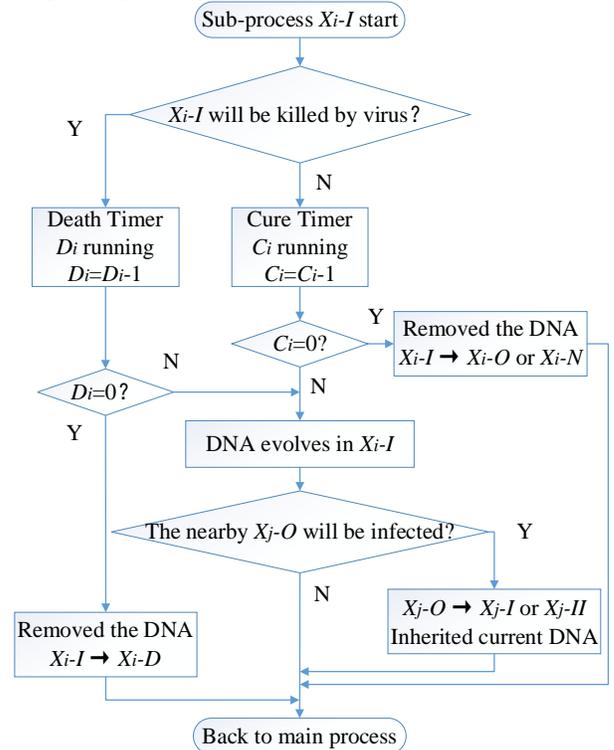

Fig. 5. Sub-flowchart of main step 5.1

The sub-process $X_i - I$ is divided into five steps.

Step 1. Determined $X_i - I$ will be killed by $V_i$ or not at the current moment. If yes, a death timer $D_i$ will be started to count the time continuously until $X_i - I$ dies. When the time of death is reached, $X_i - I$ will be transformed into $X_i - D$, and the DNA $V_i$ carried by $X_i - I$ will be destroyed. Then, go to the last step. If not, go to the next step:

$$wD = \begin{cases} wD=1 & Vle \times Dp \geq TD \\ wD=0 & Vle \times Dp < TD \end{cases} \quad (21)$$

$$Vle = \begin{cases} 100 & Gen=1 \\ 100/Gen + 40 & Gen>1 \end{cases} \quad (22)$$

$$(X_i - I) \to \begin{cases} (X_i - I) & wD=1 \,\&\&\, D_i \neq 0 \\ (X_i - I) & wD=0 \\ (X_i - D) & wD=1 \,\&\&\, D_i = 0 \end{cases} \quad (23)$$

where *Vle* is the toxicity of the virus; *Gen* represents the generation of virus; *Dp* is the death probability of the host; and *TD* is threshold.

It should be noted that if $X_i - I$ can be killed by $V_i$, $X_i - I$ will have no chance to self-cure.

Step 2. If the infected will not be killed, it means it will self-cure. It is also necessary to start a cure timer $C_i$ to continuously time until $X_i - I$ self-cure. If the self-cure time is reached, $X_i - I$ will be transformed to $X_i - O$ or $X_i - N$ with the probability $tran \in (0,1)$. This algorithm defines that the probability of transforming to immune type after self-cure is 0.3, and the DNA carried by it will be deleted. Then, go to the last step. Otherwise, go to the next step:

$$(X_i - I) \to \begin{cases} (X_i - I) & C_i \neq 0 \\ (X_i - N) & C_i = 0 \,\&\&\, tran > 0.7 \\ (X_i - O) & C_i = 0 \,\&\&\, tran \leq 0.7 \end{cases} \quad (24)$$

Step 3. DNA $V_i$ realizes the evolution process in $X_i - I$ through continuous random changes of bases, thus gradually reducing the Similarity and H-measure of $V_i$. *b* means the DNA base in $V_i$, $b_i$ represents the base at the $i^{th}$ position, *eb* represents the changed sequence, $eb_i$ is the changed base at the $i^{th}$ position, and $i \in [1,n]$:

$$eb_i = \begin{cases} r('T','C','G') & b_i = 'A' \\ r('A','C','G') & b_i = 'T' \\ r('A','T','G') & b_i = 'C' \\ r('A','T','C') & b_i = 'G' \end{cases} \quad (25)$$

$$r(x,y,z) = \begin{cases} x & 0 < ra \leq 1/3 \\ y & 1/3 < ra \leq 2/3 \\ z & 2/3 < ra \leq 1 \end{cases} \quad (26)$$

where $ra \in (0,1]$. After changing the base, we evaluate the effect of Similarity and H-measure before and after the change. If they become better, we retain the changed DNA sequence; if they become worse, we return to the original sequence:

$$b = \begin{cases} b & G(b) < G(eb) \\ eb & G(b) \geq G(eb) \end{cases} \quad (27)$$

$$G(a) = f_{Similarity}(a) + f_{H-measure}(a) \quad (28)$$

Step 4. Determined $X_i - I$ can infect $X_j - O$ within its effective spread range or not:

$$inf = (\omega_1 \tanh(S(V_i)/60) + \omega_2 Dt)/3 \quad (29)$$

$$Dt = \begin{cases} 1 & d < 0.5 \\ 0 & d > 3 \\ 1.2 - 0.4d & 0.5 \leq d \leq 3 \end{cases} \quad (30)$$

where *inf* is the probability of infection, $S(V_i)$ is the infectiousness of the DNA $V_i$, and *Dt* is the impact index of the distance between $X_i - I$ and $X_j - O$. $\omega_1$ and $\omega_2$ are the weights $\omega_1 + \omega_2 = 1$, and $\omega_1, \omega_2 \in [0,1]$. *d* is the linear distance between $X_i - I$ and $X_j - O$.

In the case of infection, $X_j - O$ will be transformed to $X_j - I$ or $X_j - II$, and then all information of the current DNA $V_i$ will be inherited:

$$(X_j - O) \to \begin{cases} (X_j - I) & rj < 0.7 \\ (X_j - II) & rj \geq 0.7 \end{cases} \quad (31)$$

where $rj \in [0,1]$. If the hosts around $X_i - I$ are not infected, then go to the last step.

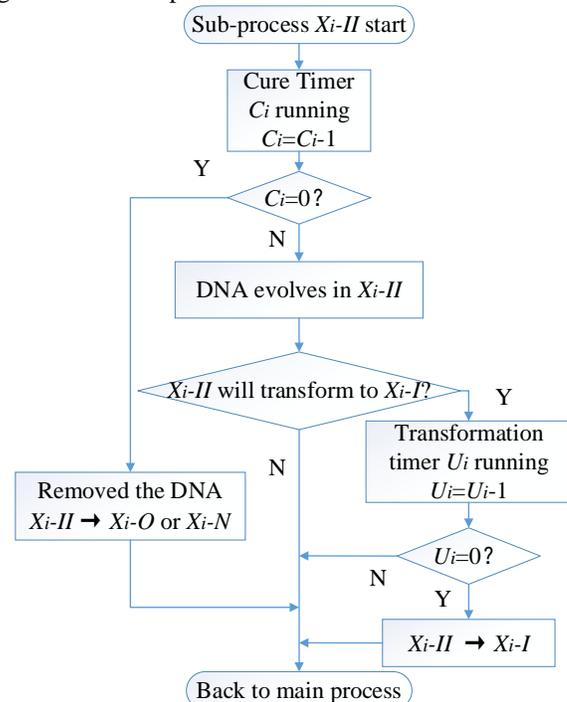

Fig. 6. Sub-flowchart of main step 5.2

Step 5. Carried all the changed parameters and returned to the main process.

Fig. 6 shows another sub-process, which is the operation for $X_i - II$ during the epidemic period, corresponding to step 5.2 in the main algorithm.

The sub-process $X_i - II$ is divided into four steps.

Step 1. $X_i - II$ will not die, so directly start a cure timer $C_i$, and keep timing until $X_i - II$ self-cure. When the time is reached, DNA $V_i$ should be deleted and transformed $X_i - II$ to $X_i - O$ or $X_i - N$, same as (24). Then, go to the last step. Otherwise, go to the next step.

Step 2. DNA $V_i$ evolves in $X_i - II$, same as (25)-(28).

Step 3. If $X_i - II$ can be transformed to $X_i - I$, then start a transformation timer $U_i$ and time until $X_i - II$ is transformed to $X_i - I$. Otherwise, go to the last step:

$$(X_i - II) \to \begin{cases} (X_i - II) & U_i \neq 0 \\ (X_i - I) & U_i = 0 \end{cases} \quad (32)$$

Step 4. Carried all the changed parameters and returned to the main process.

It should be noted that the timers $D_i$, $C_i$ and $U_i$ are the attributes of the host $X_i$ itself, which depend on its immunity and are determined at the initialization stage.

## IV. RESULTS, ANALYSIS, AND VERIFICATION

In this section, we first introduce the main parameter settings of the SVS algorithm. Then, the simulation results are analyzed and compared. Finally, the DNA sequence we designed is evaluated and biochemical validated.

### A. Main Parameters

The main parameters of the SVS algorithm in this article are shown in Table I, and the most important iteration cycle and population size are set.

TABLE I
MAIN PARAMETERS OF SVS

| Parameter | Meaning | Value |
|---|---|---|
| Ne | Number of evolutions | 20 |
| T | Total time of the virus epidemic | 20 |
| NumVirus | Initial virus size | 20 |
| Numhost | Initial host size | 300 |

### B. Simulation Results

In this part, we select HSWOA [28], NACST [29], and DMEA [32] as the comparison group and Similarity, H-measure, Continuity, Hairpin structure, GC pair Content, Melting Temperature as the compare attributes. The results are shown in Table II.

TABLE II
DNA SEQUENCES DESIGNED BY DIFFERENT METHODS

| No. | DNA sequence 5'-3' | Similarity | H-measure | Continuity | Hairpin | GC | Tm |
|---|---|---|---|---|---|---|---|
| **SVS** | | | | | | | |
| S1 | GAGTAGCTCTCTGCATAAGC | 57 | 65 | 0 | 0 | 0.5 | 62.15 |
| S2 | AATAAGAGTCGGTTCGCTCC | 56 | 67 | 0 | 0 | 0.5 | 64.18 |
| S3 | ACTCTCGGCACGTATATCAG | 55 | 58 | 0 | 0 | 0.5 | 63.38 |
| S4 | GTTCAACTAGCAATCAGCGG | 57 | 62 | 0 | 0 | 0.5 | 63.83 |
| S5 | CGCCTTATGATGGTCAACGA | 52 | 65 | 0 | 0 | 0.5 | 64.81 |
| S6 | CCTTATAGCGTGGTCCGAAT | 54 | 65 | 0 | 0 | 0.5 | 63.99 |
| S7 | ATATCACGCGTATGTCGAGC | 53 | 72 | 0 | 0 | 0.5 | 63.69 |
| **HSWOA** | | | | | | | |
| S1 | CTCGTCTAACCTTCTTCAGC | 51 | 63 | 0 | 0 | 0.5 | 62.28 |
| S2 | CTGTGTGGAATGCAAGGATG | 48 | 64 | 0 | 0 | 0.5 | 63.82 |
| S3 | CGAGCGTAGTGTAGTCATCA | 69 | 63 | 0 | 0 | 0.5 | 63.56 |
| S4 | AGTTACAGGACACCACCGAT | 51 | 65 | 0 | 0 | 0.5 | 66.39 |
| S5 | CAGTAGCAGTCATAACGAGC | 56 | 64 | 0 | 0 | 0.5 | 62.69 |
| S6 | GCATAGCACATCGTAGCGTA | 54 | 59 | 0 | 0 | 0.5 | 64.60 |
| S7 | TGGACCTTGAGAGTGGAGAT | 50 | 62 | 0 | 0 | 0.5 | 64.44 |
| **NACST** | | | | | | | |
| S1 | CTCTTCATCCACCTCTTCTC | 50 | 39 | 0 | 0 | 0.5 | 61.38 |
| S2 | CTCTCATCTCTCCGTTCTTC | 53 | 35 | 0 | 0 | 0.5 | 61.44 |
| S3 | TATCCTGTGGTGTCCTTCCT | 48 | 41 | 0 | 0 | 0.5 | 64.46 |
| S4 | ATTCTGTTCCGTTGCGTGTC | 53 | 49 | 0 | 0 | 0.5 | 65.83 |
| S5 | TCTCTTACGTTGGTTGGCTG | 50 | 48 | 0 | 0 | 0.5 | 64.63 |
| S6 | GTATTCCAAGCGTCCGTGTT | 46 | 52 | 0 | 0 | 0.5 | 65.30 |
| S7 | AAACCTCCACCAACACACCA | 40 | 52 | **9** | 0 | 0.5 | 66.71 |
| **DMEA** | | | | | | | |
| S1 | TGAGTTGGAACTTGGCGGAA | 52 | 70 | 0 | 0 | 0.5 | 63.81 |
| S2 | CAGCATGTTAGCCAGTACGA | 55 | 60 | 0 | 0 | 0.5 | 63.15 |
| S3 | TTGAGTCCGCGTGGTTGGTC | 53 | 63 | 0 | 0 | **0.6** | 64.90 |
| S4 | AATTGACACTCTGATTCCGC | 58 | 73 | 0 | 0 | **0.45** | 63.99 |
| S5 | CATACATTGCATCAACGGCG | 53 | 67 | 0 | 0 | 0.5 | 63.57 |

| | | | | | | |
|---|---|---|---|---|---|---|
| S6 | ATACACGCACCTAGCCACAC | 50 | 59 | 0 | 0 | **0.55** | 62.41 |
| S7 | GTTCCACAACAGGTCTAATG | 53 | 61 | 0 | 3 | **0.45** | 62.48 |

### C. Data Analysis and Biochemical Verification

From the numerical results, the SVS algorithm performs best in the variance of Tm and has relatively good thermodynamic properties. The difference of Tm is small, which can make the annealing operation of the biochemical reaction easier, as shown in Table III.

TABLE III
COMPARISON OF AVERAGE AND VARIANCE OF Tm

| | **SVS** | HSWOA | NACST | DMEA |
|---|---|---|---|---|
| Average | 63.86 | 63.97 | 64.25 | 63.47 |
| Variance | **0.68** | 1.86 | 4.33 | 0.77 |

With regard to Continuity and Hairpin, SVS and HSWOA are both 0, whereas NACST is slightly insufficient for Continuity. DMEA has Hairpin structures. For GC content, except DMEA, the other methods maintain 50% content.

In terms of Similarity and H-measure, the NACST method is relatively low, and the other three methods are basically the same, as shown in Fig. 7.

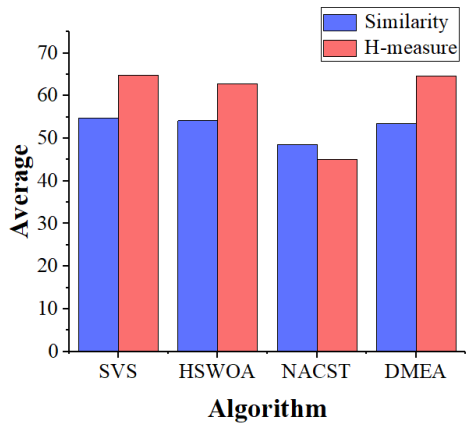

Fig. 7. Average Comparison for Similarity and H-measure

In the designed DNA sequence, NACST has a certain three-base coding, such as the sequence S1 'CTCTTCATCCACCTCTTCTC'. As the name implies, three-base coding means to design DNA with three bases, which can effectively reduce the occurrence of complementary pairing, which is the main reason why the Similarity and H-measure of NCAST in Fig. 7 are obviously superior to other methods. However, this method is only applicable to small-scale DNA sequence sets. With the increase in DNA sequence sets, there will be some problems in three-base coding, because the fewer the base types, the simpler is the way of DNA sequence base arrangement and combination. Therefore, three-base coding is effective for smaller DNA sequence sets, however, we did not use it.

Based on our actual biochemical experiment experience, we pay special attention to the use of Guanine (G). First, a large number of Guanines will form a G-quadruplex structure [43]. Second, Guanine can also play a role in quenching the fluorescence labeling [44]. Third, under a particular DNA rotation angle, Guanine and Thymine will form a misplaced pair [45]. Therefore, the use of Guanine needs care. For the pure pursuit of two DNA strands $X_i$ and $X_j$ that are not complementary, it is very effective to reduce the number of Guanines. To keep the proportion of GC base content at about 50% and reduce the number of Guanines, there will be a lot of Cytosines (C) in these sequences. Once the complement strands of $X_i$ or $X_j$ are needed, there will be a lot of Guanines in the complements, which will easily lead to the abovementioned three situations. Therefore, for universal sequence design, keeping the balance of GC base content will reduce the potential risks related to the subsequent DNA use. In the several comparison methods (Table II), the quantity of Guanine in each DNA sequence is shown in Fig. 8.

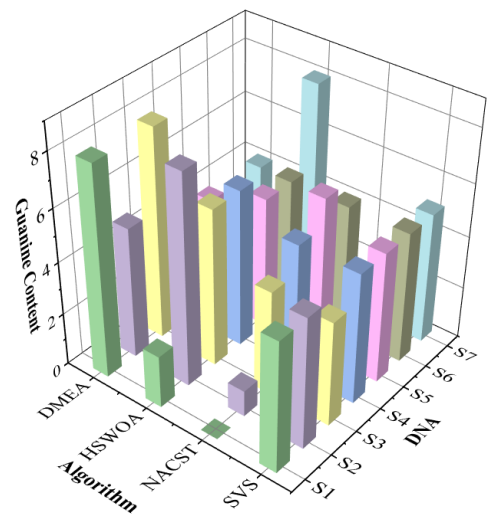

Fig. 8. Content of Guanine in Each DNA

According to the analysis in Table II and Fig. 8, in the DNA sequence designed by the SVS method, the number of Guanine is always close to 5 bases, which is more balanced and stable than that in other methods. From the perspective of numerical comparison, reducing the content of Guanine or the way of three-base coding can reduce the Similarity and H-measure of sequences. In fact, these two values represent only one type of statistical significance. The actual DNA sequence design needs to consider the content of each base. For universal coding, it is more appropriate that the content of the four bases is consistent.

The sequence design of DNA serves the molecular operations of DNA. We used experiments to verify the feasibility of the operations. We illustrate the non-complementarity of our designed sequence through polyacrylamide gel electrophoresis experiment. The electrophoresis gel is shown in Fig. 9.

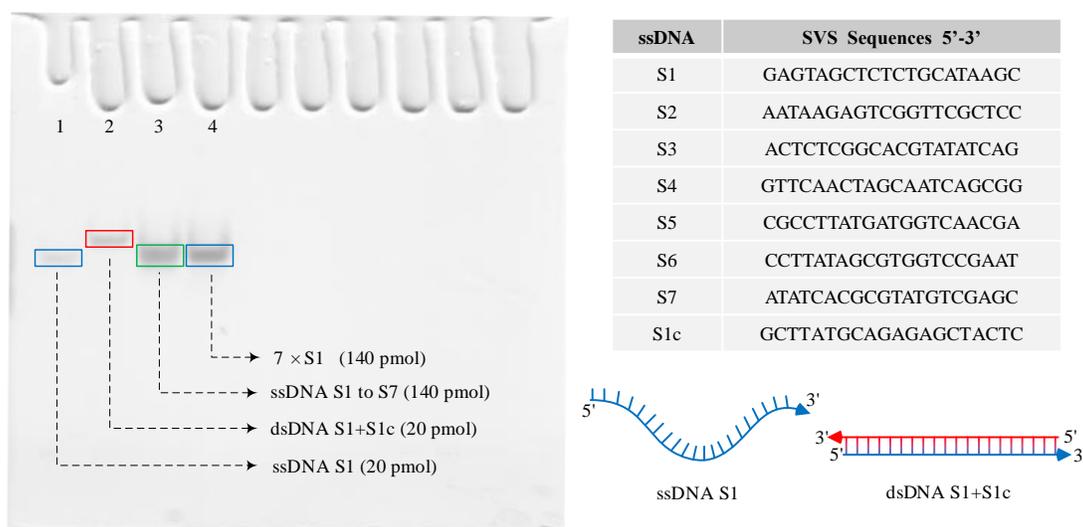

Fig. 9. Polyacrylamide Gel Electrophoresis Experiment Result for the SVS Algorithm

In Fig. 9, the substance in lanes 1 and 2 was 20 pmol ssDNA S1 and 20 pmol dsDNA (S1+S1c), respectively. All 140 pmol ssDNA from S1 to S7 was put into lane 3 and 140 pmol ssDNA S1 was put into lane 4. After mixing S1 to S7, there was still no obvious double-strand structure, which indicates that the sequence set we designed had low complementarity. DNA presents a complex attraction and repulsion states in the reaction solution and especially a large amount of base-exposed ssDNA; so, lane 3 appears slightly turbid than lane 4.

## V. CONCLUSION

DNA sequence design is a basic component of DNA computing. Designing reliable and accurate sequences is of great value for the stability of biochemical reactions. In this paper, to obtain a large number of DNA sequence sets with low complementarity probability, we proposed the SVS algorithm, which combines a variety of constraints to optimize and expand a low-complementarity DNA sequence set while also ensuring a small difference in Tm and a balanced distribution of base types. To improve the efficiency of the algorithm, we used self-cure, death, and immune mechanisms to effectively decrease the search space and increase the convergence. The final computer simulation and gel electrophoresis experiment showed that the SVS algorithm was effective in obtaining high-quality DNA sequence sets.

Although fixing the location of all hosts can strengthen the convergence of the algorithm, it also limits the global search ability. In future research, on the one hand, we will design a dynamic virus spread algorithm to increase the global search capability; on the other hand, we will explore methods and theories of non-universal DNA sequence design, including but not limited to special recognition domain sequences of enzymes, ion-bridge base mismatch sequences, and metal nanoparticle synthesis template sequences. Our algorithm can be improved in terms of global search and convergence, and it can also assist in performing biochemical reactions with special needs.

## VI. SUPPLEMENTARY MATERIALS

In the supplementary materials, we introduce the relevant buffer formula, annealing, and other conditions for the polyacrylamide gel electrophoresis experiment. We also provide basic statistical information on the virus spread under certain parameters and the epidemic dynamic diagram.

## VII. ACKNOWLEDGEMENTS

## VIII. DECLARATION OF INTEREST STATEMENT

The authors declare that they have no known competing financial interests or personal relationships that could have appeared to influence the work reported in this paper.